\documentstyle[12pt]{article}
\def \lim{\rm Lim}

\begin{document}
\baselineskip=24pt
\bibliographystyle{plain}
\title{Linear Sigma Model Linkage with Nonperturbative QCD}

\bigskip

\bigskip

\author{
L.~R.~Baboukhadia, V. Elias $^*$ and M.D. Scadron, \\ Physics Department, 
University of
Arizona, Tucson, AZ 85721 USA \\ Fax: (602) 621-4721} 
\date{}
\maketitle
\bigskip

\bigskip

\begin{abstract}
\baselineskip=24pt

Linkage is demonstrated between the quark mass, quark condensate and
coupling strength in the infrared limit of QCD and in the
quark-level linear sigma model.
\end{abstract}

\vfill


\vfill

\noindent $*$Permanent address: Department of Applied Mathematics, \\
University of Western Ontario, London N6A 5B7, CANADA \\

\newpage

\section*{1  Introduction}

\smallskip

\ \ \ \ \ Although QCD appears to be the best candidate among existing
effective-field-theoretical models for a complete picture of
the strong interactions, its non-perturbative, low-energy
behaviour is less-well-understood than its high energy behaviour.  
Consequently, it remains necessary to consider models for the strong
interactions that, on the one hand, exhibit linkage to low-energy QCD
physics, and, on the other hand, facilitate computations of key 
phenomenological
quantities.

\smallskip

Guidance in this search naturally relies upon chiral symmetry, which plays 
a
dominant role in governing low-energy phenomena in the strong
interactions.  Consequently, it is important to look for the simplest
chiral realization of QCD, which we believe is either the
Nambu-Jona-Lasinio (NJL) scheme or the Linear Sigma Model (L$\sigma$M).

\smallskip

In this context, we consider a quark-level L$\sigma$M where
elementary fermions are treated as quarks and are combined with
elementary $\vec{\pi}$ and $\sigma$ mesons in a chiral-invariant way. 
Recent evidence for the occurrence of a $\sigma$-meson in $\pi \pi$
scattering [1,2] provides additional motivation for examining the
linkage between the linear sigma model and QCD.  Moreover, some
analyses suggest that this scalar meson has a mass at or near twice
the dynamical mass anticipated for light quarks [2], consistent with
L$\sigma$M and NJL expectations [3,4].

\smallskip

In what follows, we examine the correspondence 
between nonperturbative QCD at freeze-out energy scales and the
L$\sigma$M, based upon the equivalence (or near
equivalence) of quark masses, quark-antiquark condensates, and coupling
strengths.

\smallskip

In Section 2 we work out in detail the relationship between the dynamically
generated quark mass and the nonperturbative quark condensate in QCD.  Then
in Section 3 we compute these nonperturbative quantities in the
quark-level linear sigma model (L$\sigma$M).  Next in Section 4 we show how
the coupling constant strength associated with chiral symmetry breakdown 
in QCD corresponds
to the anticipated meson-quark coupling strength in the L$\sigma$M. 
Finally, in Section 5 we discuss how a reappraisal of the $\pi^\circ
\rightarrow \gamma\gamma$ decay process provides insight into 
effective theories of low-energy QCD, such as the L$\sigma$M, that contain
constituent-mass quarks.

\bigskip

\section*{2  Dynamical Quark Mass and Condensate in QCD}

\smallskip

\ \ \ \ Within QCD, the relationship of quark-antiquark condensates 
to observable physics
is even more tenuous (or, more indirect)
than that of quark masses.  Consequently, there is some value in
reviewing the relationship between these two order-parameters of
chiral noninvariance.  The existence of a nonperturbative quark-antiquark 
vacuum condensate
is fundamental to the dynamical generation of a
nonperturbative quark mass.  To see this, consider the limit in which
no Lagrangian quark mass appears.  In the Wick-Dyson expansion of the
time-ordered product of fermion-antifermion fields,
\begin{eqnarray}
T \psi_i (x) \psi_j(y) & = & <O_p \mid T \psi_i (x) \overline\psi_j (y)
\mid O_p > \nonumber \\
& + & : \psi_i (x) \overline\psi_j (y): , 
\end{eqnarray}
the expectation value of the time-ordered product between
purely-perturbative vacuum states $\mid O_p >$ is the massless
free-fermion propagator
\begin{eqnarray}
< O_p \mid T \psi_i^\alpha (x) \overline\psi_j^\beta (y) \mid O_p >
\nonumber \\
 =  \int \frac{d^4k}{(2 \pi)^4} i \delta^{\alpha\beta}
\frac{\gamma_{ij}^\mu k_\mu}{k^2} e^{-ik \cdot (x-y)},
\end{eqnarray}
where \{$\alpha,\beta$\} are internal symmetry indices (e.g. colour)
and \{$i, j$\} are Dirac spinorial indices. 
Let us assume, however, that we are calculating in the presence of a
vacuum $\mid \Omega >$ whose nonperturbative content admits the existence 
of a
fermion-antifermion condensate:
\begin{equation}
< \overline\psi \psi > = - \stackrel{\lim}{_{x \rightarrow y}} <\Omega 
\mid :
\psi_i^\alpha (x) \overline\psi_i^\alpha (y): \mid \Omega > .
\end{equation}

\smallskip

The vacuum $\mid \Omega >$ is clearly nonperturbative in character,
since normal-ordered fields necessarily annihilate the
purely-perturbative vacuum $\mid O_p >$.  The full fermion propagator
is obtained by taking the expectation value of (1) between
nonperturbative vacuum states $\mid \Omega >$,
\begin{eqnarray}
< \Omega \mid T \psi_i^\alpha (x) \overline\psi_j^\beta (y) \mid
\Omega > = < O_p \mid T \psi_i^\alpha (x) \overline\psi_j^\beta (y)
\mid O_p > \nonumber \\
 + < \Omega \mid : \psi_i^\alpha (x) \overline\psi_j^\beta (y) :
\mid \Omega > .
\end{eqnarray}

\smallskip

It is evident from (4) that the existence of a quark-antiquark
condensate, as defined by (3), implies a difference between the full
fermion propagator \hfill \\
$< \Omega \mid T \psi_i^\alpha(x)\overline\psi_j^\beta(y) \mid \Omega>$ 
and the massless free
fermion propagator (2). In Landau gauge, in which leading-order
quark-antiquark-condensate contributions to the spinorial component
of the fermion self-energy are seen to vanish [5],this difference 
can be ascribed to a dynamical mass function in the
full fermion Landau-gauge propagator,
\begin{eqnarray}
& < & \Omega \mid T \psi_i^\alpha (x) \overline\psi_j^\beta (y) \mid
\Omega > \nonumber \\
& = & \int \frac{d^4k}{(2\pi)^4} i \delta^{\alpha\beta}
\frac{[\gamma_{ij}^\mu k_\mu + M(k^2) \delta_{ij}]}{[k^2 - M^2(k^2)]}
e^{-ik \cdot(x-y)}
\end{eqnarray}
One can rearrange (4) to relate this dynamical mass function
directly to the quark-antiquark vacuum condensate [5,6]
\begin{eqnarray}
- <\overline\psi \psi > &=& \stackrel{\lim}{_{x \rightarrow y}} Tr < \Omega
\mid : \psi_i^\alpha(x) \overline\psi_i^\alpha (y) : \mid \Omega >
\nonumber \\
& = & N i \int \frac{d^4k}{(2\pi)^4} Tr \left[ \frac{\gamma \cdot k +
M(k^2)}{k^2 - M^2(k^2) + i \epsilon} - \frac{\gamma \cdot k}{k^2 + i
\epsilon} \right] \nonumber \\
& = & 4 N i \int \frac{d^4k}{(2\pi)^4} \left[ \frac{M(k^2)}{k^2 - M^2(k^2)
+ i \epsilon} \right] \, ,
\end{eqnarray}
where $N \equiv \delta^{\alpha\alpha}$, the sum of internal symmetry
indices [e.g. colour number $N_c$].

\smallskip

Equation (6) can be understood diagrammatically as the equivalence of the
quark-antiquark condensate to a quark propagator-loop over the full
fermion propagator [Fig 1].  This equation, however, demonstrates
that the existence of a dynamical mass function $M(k^2)$ is
necessarily linked to the nonperturbative content of the vacuum $\mid
\Omega >$, as parametrised by the quark-antiquark condensate defined
in (3).

\smallskip

It is evident from (6) that the dynamical quark mass scale and the
scale of the quark-antiquark condensate are tightly linked.  We will
assume that $M(k^2)$ is asymptotically even as $k^2$ flips from
timelike to spacelike values [5].  

If $m_{dyn}$ is identified with the pole [4,7,8] of the fermion
propagator in (5),
\begin{equation}
m_{dyn} = M(m^2_{dyn}),
\end{equation}
and if $M(k^2)$ is analytic, \footnote{Controllable low-energy behaviour 
and the absense of IR-divergences in the mass function $M(p^2)$
is suggested by the leading-order ``freezing out'' of $M(p^2)$ at
$| p^2 |$ less than $m^2_{dyn}$ [8,9].} then the left-hand
side of the dispersion relation

\begin{equation}
Im \left( M(m^2_{dyn}) \right) = - \frac{1}{\pi} P
\int_{-\infty}^{\infty} dp^2 \frac{Re(M(p^2))}{p^2 - m^2_{dyn}},
\end{equation}
must vanish in order to ensure a quark decay-width of zero (the quark
is stable).  We assume the right-hand side of
(8) is dominated by the asymptotic $(p^2 \rightarrow \pm
\infty)$ behaviour of the integrand. If such is the case, a vanishing 
result
is possible only if the integrand of (8) is
asymptotically odd in $p^2$, in which case $Re [M(p^2)]$ must be
asymptotically {\it even} in $p^2$.

\smallskip

One approximate way of relating the dynamical quark mass to the scale
of the condensate is to identify the momentum dependence of $M(k^2)$
with the asymptotic momentum dependence anticipated from
Bethe-Salpeter and renormalization-group considerations [7]
\begin{equation}
M(k^2) = \frac{\mu^2 M(\mu^2)}{\mid k^2 \mid} 
\left( \frac{ \ln ( \mid
k^2 \mid / \Lambda_{QCD}^2)}{ \ln (\mu^2 / \Lambda_{QCD}^2)}
\right)^{d-1} .
\end{equation}
In (9), $\mu^2$ is a renormalization scale,and the
anomalous-dimension factor $d$ is given by
\begin{equation}
d = 9(N_c^2 - 1) / [2 N_c (11 N_c - 2 N_f)]
\end{equation}
for an SU(N$_c$) internal symmetry group with $N_f$ fermion flavours.  
Strictly speaking, (9)
describes asymptotic behaviour of M for $\mid k \mid^2 >>
\Lambda_{QCD}^2$ in the Euclidean regime.  The absolute value signs
in (9) are to ensure that $M(k^2)$ is asymptotically even, as
discussed above.

\smallskip

We now seek to relate the renormalized condensate $<\overline\psi \psi
>_\mu \sim [ln(\mu^2 / \Lambda^2)]^d$ to the dynamical mass
function (9).  To extract the leading-logarithmic [i.e.,
renormalization-group (RG)] dependence of
the condensate,  we substitute (9) into (6), with UV cut off $\mu$ and 
$IR$ cut off
$\Lambda_{QCD}$, and find that [5]
\begin{equation}
- < \overline\psi \psi >_\mu = \frac{4 N_c Ci}{(2 \pi)^4} \int
_{\Lambda_{QCD}}^\mu  
\frac{d^4k [ \, \ln [ \mid k \mid^2 / \Lambda_{QCD}^2]]^{d-1}}{\mid k^2 
\mid (k^2 -
M^2(k^2))},
\end{equation}
where the constant $C$ is given by
\begin{equation}
C \equiv \mu^2 M(\mu^2) \left[ \ln (\mu^2 / \Lambda_{QCD}^2)
\right]^{1-d}.
\end{equation}
For $\mu^2$ to be sufficiently large to justify neglect of $M^2(k^2)$ in 
the
integrand denominator, one can perform a Wick rotation $d^4k = i
\pi^2 k_E^2 dk_E^2$ and evaluate (11) explicitly:\footnote{Arc
contributions of a finite but large radius occurring after Wick
rotation have been neglected, as they vanish in the large $\mu$ limit
and, hence, do not contribute to the leading-log expression for
$<\bar{\psi} \psi >$.}

\begin{eqnarray}
- < \overline\psi \psi >_\mu & = & \frac{N_cC}{4 \pi^2 d} \left[ \ln
\left( \frac{\mu^2}{\Lambda_{QCD}^2} \right) \right]^d \nonumber \\
& = & \frac{N_c \mu^2 \, M(\mu^2)}{4 \pi^2d}  \ln \left(
\frac{\mu^2}{\Lambda_{QCD}^2} \right) .
\end{eqnarray}
One can eliminate the (leading-)logarithm by utilizing the one-loop RG 
expression
for the running QCD coupling constant
\begin{eqnarray*}
\alpha_s(\mu^2) = \pi d'/[\ln (\mu^2 / \Lambda_{QCD}^2)],
\end{eqnarray*}
\begin{equation}
d' \equiv 12 / (11 N_c - 2N_f),
\end{equation}
to find from (13) that
\begin{equation}
- < \overline\psi \psi >_\mu = \frac{ N_cd' \mu^2 M(\mu^2)}{4 \pi d
\, \alpha_s (\mu^2)} = \frac{2 N^2_c \mu^2 M(\mu^2)}{3 \pi(N^2_c - 1)
\alpha_s (\mu^2)}.
\end{equation}

\smallskip

We note that as the scale $\mu$ is varied, $\mu^2 M(\mu^2) \sim
[\ln (\mu^2 / \Lambda_{/QCD}^2)]^{d-1}$ and $\alpha_s (\mu^2) \sim
[\ln (\mu^2) / \Lambda_{QCD}^2)]^{-1}$.  Thus we confirm that $<
\overline\psi \psi >_\mu \sim [\ln (\mu^2 / \Lambda_{QCD}^2)]^d$,
consistent with the RG invariance of $m
\overline\psi \psi$ Lagrangian mass terms [the current quark mass
runs as $m(\mu^2)
\sim \left( \ln (\mu^2 / \Lambda_{QCD}^2) \right)^{-d}$]. 
Consequently, (15) must be regarded as a relationship between the
RG quantities $<\overline\psi \psi>_\mu$ and $M(\mu^2)$, consistent
with the large-$\mu$ assumptions employed in evaluating (11) and (13).

\smallskip

The result (15) is in fact consistent with that obtained directly
from the quark-antiquark condensate contribution (Fig. 2) to the
Landau-gauge self-energy, as determined [10] using operator-product
expansion methods appropriate for spacelike $(k^2 < 0)$ momenta:
\begin{equation}
M(k^2) = \frac{3(N_c^2 - 1)}{2 N_c^2} \frac{\pi \alpha_s}{k^2} <
\overline\psi \psi > .
\end{equation}
The use of Landau gauge decouples leading-order wave-function 
renormalization
effects from the quark-antiquark condensate [5], and can be motivated 
to decouple dynamical-mass-function contributions to vertex functions
as well [11].  Renormalization-group improvement
of (16) leads to
\begin{equation}
M(k^2) = \frac{3(N_c^2 - 1)}{2 N_c^2}  \frac{ \pi\alpha_s ( \mid k^2
\mid )}{ \mid k^2 \mid} \mid < \overline \psi \psi >_k \mid,
\end{equation}
where even behaviour in $k^2$ [consistent with the known positivity of
$M(k^2)$ when $k^2 < 0$] has been imposed for reasons given earlier. 
Upon algebraic rearrangement of (17) with $k \rightarrow \mu$, eqs.
(17) and (15) are seen to be equivalent - the loop integral [Fig. 1]
over the RG-improved Bethe-Salpeter dynamical mass function and the
Fig 2 contribution to the quark self-energy generate equivalent
leading-order expressions.  If one chooses $k^2 = m^2_{dyn}$ in (17)
and utilizes the initial condition (7), one finds that\footnote{In
(15), we note that $d = d'$ when $N_c = 3$.  We have chosen to keep
$N_c$ arbitrary in order to demonstrate that the equivalence of the
Fig. 1 approach to (15) and the Fig. 2 approach to (17) is {\it not}
a consequence of this coincidental equality.  We are grateful to R.
Tarrach for making us aware of this concern some time ago.}

\begin{equation}
- < \bar{\psi} \psi > _{m_{dyn}} = \frac{2 N^2_c m^3_{dyn}}{3(N^2_c -
1) \pi \alpha_s (m^2_{dyn})}.
\end{equation}

\smallskip

In Section 4 we argue that
$\alpha_s(m_{dyn}^2)$ can be identified with the critical value
$\pi/4$ associated with the onset of dynamical mass generation
through the Schwinger-Dyson equation for the quark propagator
[12,13].  With $N_c = 3,
\, \, \alpha_s = \pi/4$, we see from (18) that 
\begin{equation}
- < \overline\psi \psi >_{m_{dyn}} = \frac{3m_{dyn}^3}{\pi^2},
\end{equation}
in which case 
\begin{equation}
- < \overline\psi \psi>_{1 GeV} = \frac{3m_{dyn}^3}{\pi^2} \left[ \ln
\left( \frac{1 GeV^2}{m_{dyn}^2} \right) \right]^{4/9}
\end{equation}
at the 1 GeV momentum scale characterizing QCD sum-rule estimates.
Thus, a dynamical quark mass of 320 MeV corresponds to a 1 GeV
quark-antiquark condensate $< - \overline\psi \psi >_{1 GeV} = (243
\, MeV)^3$, 
consistent with QCD sum-rule estimates [14].

\bigskip

\section*{3  Condensate and Quark Mass in the Linear Sigma
Model}

\smallskip

\ \ \ \ \ The linear sigma model [15] treated as an effective theory
for the interaction of constituent quarks and mesons [16,17] is
controlled by the (chiral) quark level Goldberger-Treiman (GT)
relation $g = m_q / f_{\pi}$ ensuring $\partial \cdot
\vec A = 0$ in the
chiral limit (CL).  Thus a quark mass of 320 MeV (as in the previous
section) and the CL of the pion decay constant, \footnote{One can show 
that $1 -
f_\pi^{CL} / f_\pi = m_\pi^2/8\pi^2f_\pi^2 \approx 0.03$ or $f_\pi^{CL}
\approx$90 MeV for the physical $f_\pi \approx$ 93 MeV [18].} $f_{\pi} 
\approx 90
GeV$  correspond to a $\pi q \overline q$ coupling constant $g
\approx 3.6$, a value 
compatible with the observed [19] pion-nucleon coupling
constant $g_{\pi N N} = 3gg_A$ between $13.0$ and $13.5$. \footnote{A
$\chi^2$-minimization corresponding to $g_{\pi N N} = 13.145 \pm 0.072$
is presented in the most recent paper in ref. [19].}  Elsewhere [4,16], 
the 
quark-level linear sigma model (L$\sigma$M) $\pi q \overline q$ coupling 
has been {\it predicted} to be
$g = 2 \pi / \sqrt{3} \approx 3.63$ for $N_c = 3$.  This prediction
corresponds to a GT quark mass $m_q = 2 \pi f_{\pi} / \sqrt{3} \cong
326 MeV$, consistent with $m_{dyn} \approx M_N / 3$ phenomenological
expectations.

\smallskip

This ``hard'' quark mass can be used via (6) to determine the
linear-sigma model quark-loop analogue of the quark-antiquark
condensate for $N_c = 3$:
\begin{equation}
- < \overline\psi \psi>_{L \sigma M} = \frac{ 12 i}{(2 \pi)^4}
\int_0^\Lambda \frac{ d^4k}{k^2 - m_q^2} m_q .
\end{equation}
In the L$\sigma$M, the ultraviolet cutoff $\Lambda$ is determined by
the logarithmically divergent quark loop diagram for $f_{\pi}$ [Fig.
3]
\begin{equation}
if_\pi q_\mu = 12g m_q q_\mu \int_0^\Lambda \frac{ d^4k}{(2 \pi)^4
[k^2 - m_q^2]^2} .
\end{equation}
Since $f_\pi = m_q / g$, eq. (22) leads to the gap equation
\begin{eqnarray}
1 & = & -12 i g^2 \int_0^{\Lambda} d^4 p (2 \pi)^{-4} (p^2 -
m_q^2)^{-2} \nonumber \\
& = & \frac{ 3g^2}{4 \pi^2} \left[ \ln \frac{ \Lambda^2 +
m_q^2}{m_q^2} - \frac{\Lambda^2}{\Lambda^2 + m_q^2} \right].
\end{eqnarray}
If $g = 2 \pi / \sqrt{3}$, the square-bracketed expression in (23) is
unity for $\Lambda = 2.30 m_q$.  With this value of $\Lambda$, one finds 
from (21) that
\begin{eqnarray}
- < \overline\psi \psi >_{L \sigma M} & = & \frac{3m_q^3}{4 \pi^2}
\left[ \frac{\Lambda^2}{m_q^2} - \ln \left( \frac{\Lambda^2}{m_q^2} +
1 \right) \right] \nonumber \\
&\cong&(209 MeV)^3
\end{eqnarray} 
for $m_q = 326 MeV$.

\smallskip

It should be noted that the corresponding $\Lambda$ value (750 MeV)
is both somewhat above the L$\sigma$M scalar meson mass $m_\sigma = 2m_q$ 
and
somewhat below the $\rho$-mass, consistent with phenomenological 
expectations. 
The L$\sigma$M describes an effective theory for low-energy QCD in which
the $\sigma$-meson is fundamental but the $\rho$-meson is composite.  The
condensate value in (24) can be compared with the corresponding QCD
estimate from the previous section at momentum scale $\Lambda$.  We
suggest here that it may be more appropriate to associate L$\sigma$M
parameters with their corresponding QCD parameters in the infrared
limit; i.~e. at their low energy ``freeze-out'' values.  In the next
section, we discuss the freezing out of the QCD
coupling constant at the critical value $\alpha_s = \pi/4$. 
As remarked earlier, a similar freezing out of the dynamical quark mass 
function $M(k^2)$ at the critical value $m_{dyn}$ has been
demonstrated
both by plane wave [8] and coordinate space [9] methods.
It is evident from
(15) and (18) that the latter equation defines $< \overline\psi \psi
>$ in terms of ``frozen out'' values for $\alpha_s$ and $M(k^2)$.  Thus
(19) [equivalent to (18) with explicit use of $\alpha_s (m_{dyn}) =
\pi/4$] represents the magnitude of $< \overline\psi \psi>$ at its
low-energy ``freeze-out''.  Comparison of (19) to (24) is suggestive
of near-equality between the L$\sigma$M quark mass (326 MeV) and the
QCD dynamical mass (311 MeV) that would generate equivalence between
the L$\sigma$M condensate $< \overline\psi \psi>_{L \sigma M}$ and
the ``freeze-out'' QCD condensate $<\overline\psi \psi>_{m_{dyn}}$. 
Alternatively, if $m_{dyn} = 326 MeV$ (the L$\sigma$M value), we find
from (19) that $< - \overline\psi \psi>_{m_{dyn}} = 3m_{dyn}^3 / \pi =
(219 \, MeV)^3$, a value quite comparable to (24). 

\bigskip

\section*{4  Correspondence Between L$\sigma$M and QCD Coupling Strengths}
      
\smallskip

\ \ \ \ \ Thus far we have argued that the link between ``frozen-out'' QCD 
and
the L$\sigma$M is in the near-equivalence of their order-parameters of 
chiral-noninvariance:
\begin{equation}
m_{dyn} = m_q \, \, \, , \, \, \, <\overline \psi \psi>_{m_{dyn}} \approx
<\overline \psi \psi>_{L \sigma M} .
\end{equation}  
Correspondence between QCD and L$\sigma$M 
coupling strengths can be obtained from the quark mass gap $\sigma$-tadpole
graph of Fig. 4, which leads to the L$\sigma$M relation [4]
\begin{equation}
m_q = \frac{ g_{\sigma qq}^2}{-m_\sigma^2} < \overline \psi \psi>_{L \sigma
M},
\end{equation}
for zero momentum transferred to the vacuum.  On the other hand, we
see from (18) (with $N_c = 3$) that [5,20] 
\begin{equation}
< - \overline \psi \psi>_{m_{dyn}} = (3 / 4 \pi \alpha_s)m_{dyn}^3 ,
\end{equation}
where $\alpha_s$ is assumed to be at its infrared ``frozen-out'' value. 
We use (25) to justify substitution of (27) into (26) and find that
\begin{equation}
m_\sigma^2 / m_q^2 \approx 3 g_{\sigma qq}^2 / 4 \pi \alpha_s .
\end{equation}

\smallskip

Since one knows that $m_\sigma = 2m_q$ in the Nambu Jona-Lasinio model
[3] as well as in
the quark-level
L$\sigma$M [4,16] in the chiral limit, (28)
then implies that
\begin{equation}
g_{\sigma qq}^2 / 4 \pi \approx (4/3) \alpha_s \equiv \alpha_s^{eff} .
\end{equation}
Eqs. (25) and (29) together constitute the linkage between L$\sigma$M and
infrared QCD parameters.

\smallskip

Now we proceed to examine separately the scales of the L$\sigma$M and
QCD couplings related by (29).  The QCD coupling strength
$\alpha_s = g_s^2 / 4 \pi$ runs according to (14) in one-loop order.
The ALEPH Collaboration [21] value of $\alpha_s(M_Z^2) = 0.122 \pm
0.007$ can be used [22] to obtain $\alpha_s(2.5 \, GeV^2) = 0.375 \pm
0.07$ at the threshold of the three-flavour region.  Substituting
this latter coupling back into (14) with $d^{\prime} = 4/9$ for three quark
flavours, one finds $\Lambda_{QCD} \approx 246 MeV$.  This
three-flavour QCD cutoff in turn generates $\alpha_s(1 GeV^2) \approx 0.5$,
which is the usual QCD coupling one expects at the $\phi$(1020) scale
[23]. \footnote{Our own calculation, using somewhat smaller PDG [1]
world average for $\alpha_s (m_z^2) [0.118 \pm 0.03]$ evolved down to
$2.5 \; GeV^2$ via the 3-loop $\overline{MS} \; \beta$-function
[1], yields the range $\alpha_s = 0.338_{-0.027}^{+0.030}$, where the 
uncertainty includes the PDG
uncertainty in the five-flavour threshold [$ 4.1 \; GeV \leq m_b \leq 4.5
\; GeV$].  The corresponding range of $(1-$loop$) \; \Lambda$ in (14) is
between 167 and 237 MeV, consistent with $\alpha_s(1 GeV^2)$ between
0.4 and 0.5. }

\smallskip

We focus here on the value of $\alpha_s$ at the NJL - L$\sigma$M
scalar mass $m_\sigma = 2m_q \approx 650 MeV$, which for
$\Lambda_{QCD} = 246 MeV$ is $\alpha_s [(650 MeV)^2] \cong 0.72$,
using (14) again.  This
value is to be compared with Mattingly and Stevenson's [24] extension
of QCD to the infrared region via ``minimal sensitivity'', leading to
a freezing out of $\alpha_s$ at the value $\alpha_s [(2m_q)^2] / \pi =
0.26$, [$\alpha_s$ = 0.82].  The
relevant issue here, however, is not so much the location of the
freeze-out momentum scale, \footnote{Mattingly and
Stevenson in [24] identify $m_q$ with the current quark mass.} but rather 
the near equivalence
of their
numerical determination of the frozen coupling to the value
$\pi/4(=0.785)$ anticipated from chiral symmetry breaking.  That is,
freezeout is seen to occur at or near $\alpha_s = \pi / 4$.  

\smallskip

Theoretical justification for this value follows from the
``supercriticality'' [25, 26] occurring when $\alpha_s^{eff} = (4/3)
\alpha_s = \pi / 3 (\approx 1)$.  At this strength, the underlying
Dirac equation breaks down, similar to the perturbative breakdown
occurring in large-Z atoms when Z$\alpha \rightarrow 1$. Such spontaneous 
breakdown in QCD
is presumably characterized by a nonvanishing quark condensate $<
\overline q q>_0 \neq 0$.  Stated another way, when $\alpha_s^{eff} =
\pi / 3$, Bethe-Salpeter dynamics is at a singular point [25] for
nonperturbative QCD, or at an ``ultraviolet fixed point'' for abelian,
quenched planar QED [27].  Near this singular point the $0^-$, $\overline
q q$ Bethe-Salpeter bound state wave function in Landau gauge has the
asymptotic form
\begin{equation}
\chi(p^2) \sim (p^2)^{-(3 \pm \sqrt{1 - 3 \alpha_s^{eff} / \pi)} / 2} ,
\end{equation}
which of course suggests $\alpha_s^{eff} \rightarrow \pi/3$ at
breakdown.  Alternatively, the Schwinger-Dyson approach of
Higashijima [12] identifies the onset of chiral symmetry breakdown
when $\lambda \equiv (3 / \pi) C_2 (R) \alpha_s > 1$, confirming the
criticality at $\pi/3$ of $\alpha_s^{eff} = C_2(R) \alpha_s$.

\smallskip

A more qualitative understanding of $\alpha_s^{eff} \sim 1$ at
freezeout is based on the uncertainty principle $p \sim 1/r$ for a
relativistic tightly bound $\overline q q$ pion with $m_\pi = 0$
[12].  The total pion energy for confining QCD potential $V = -
\alpha_{eff} / r$ is then
\begin{equation}
E_\pi = KE + V = p - \alpha_s^{eff} / r \sim p(1 - \alpha_s^{eff}).
\end{equation}
Then $E_\pi > 0$ is physical for $\alpha_s^{eff} < 1$, with
$\alpha_s^{eff} = 1$ corresponding to $E_\pi$ becoming unphysical - a
signal of the quarks condensing.  Further confirmation of this
relativistic picture of the pion is that this tight binding (with
$E_\pi \rightarrow 0$) corresponds to the two (constituent) quarks in
the $\overline q q$ pion fusing together with net (L$\sigma$M) pion
charge radius [28]
\begin{equation}
r_\pi = \sqrt{3} / 2 \pi f_\pi = 1 / m_q \approx 0.61 \, fm ,
\end{equation}
in close agreement with the measured $r_\pi$.

\smallskip

Finally, we note from (29) that the value
$\alpha_s^{eff} = \pi / 3$ implies that the L$\sigma$M coupling
constant is $g_{\sigma qq} = 2 \pi / \sqrt{3}$.  This is precisely the 
value
anticipated ($g \equiv g_{\sigma qq}$ in Section 3) from L$\sigma$M 
phenomenology
[4] (and
predicted in [16]), providing 
a startling confirmation of the correspondence between QCD in the infrared 
limit
and the quark-level L$\sigma$M.
\bigskip

\section* {5  Discussion: $\pi^\circ \rightarrow \gamma\gamma$ and 
Effective Theories with
Constituent Quark Masses}

\smallskip

\ \ \ \ \ There is presently no direct derivation from the QCD lagrangian 
of either the linear sigma
model or of {\it any} phenomenological model for low energy QCD with 
constituent-mass
quarks.  In the absence of such a derivation, the decay $\pi^\circ 
\rightarrow \gamma\gamma$
is of genuine value in discussing the viability of low energy 
approximations to QCD.  In this
context, it should be remembered that the axial anomaly as well as partial 
conservation of the
axial-vector current (PCAC) were originally developed from arguments based 
upon the linear
sigma model.

      In the linear sigma model, one explicitly calculates a "Steinberger"
pseudo-scalar-vector-
vector (PVV) quark triangle graph in order to obtain the correct 
$\pi^\circ \rightarrow
\gamma\gamma$ decay amplitude [29,30]. There are no meson-loop 
contributions to the
amplitude because there is no triple pion coupling in the linear sigma 
model (such couplings are
forbidden because of parity, Lorentz invariance, and G-parity 
considerations).  The PVV loop
calculation is, of course, easily seen to be driven by the axial anomaly
[31] if one utilizes the right
hand side of the axial Ward identity using constituent quark masses:

\begin{equation}
2 m_u \overline{u} \gamma_5 u - 2m_d \overline {d} \gamma_5 d = 
\partial_\mu J_5^{3 \mu}
-
 \alpha \tilde{F} \cdot F / 2 \pi .
\end{equation}

When the quark mass is sufficiently large compared to the pion mass, the 
contribution of
$\partial_\mu J_5^{3 \mu}$ is zero by virtue of the Sutherland-Veltman 
theorem
[32], and
the PVV loop result is well-known to be driven entirely by the anomaly 
term.  In fact, the linear
sigma model PVV decay amplitude is remarkably insensitive to the quark 
mass, remaining within
(now small) experimental error for quark masses as small as 220
MeV [33] . 

      These results are entirely consistent with those obtained via PCAC 
from the QCD
lagrangian, in which only very small (current-) quark masses are expected 
to appear. Using
PCAC, one can transform the $\pi^\circ \rightarrow \gamma\gamma$ decay 
amplitude into the
divergence of the axial-vector vertex of a AVV quark loop. One then 
utilizes the axial Ward
identity in the chiral $m_{u,d} \rightarrow 0$ limit,

\begin{equation}
[f_\pi m_\pi^2 \pi^\circ = ] \, \, \partial_\mu J_5^{3 \mu} = \alpha
\tilde{F} \cdot F / 2 \pi ,
\end{equation}
to relate the divergence of the axial-vector vertex {\it directly} to the 
anomaly term
[34]. 

      This striking compatibility between the $\pi^\circ \rightarrow 
\gamma\gamma$ decay
amplitude appropriate for QCD-lagrangian quark masses [and obtained via 
PCAC] and the
$\pi^\circ \rightarrow \gamma\gamma$  decay amplitude for constituent 
quark masses within a
linear sigma model context {\it is no longer evident} in the absence of 
the pion-quark couplings
characterizing
the linear sigma model. In the absence of such couplings, any 
phenomenological model involving
constituent mass quarks must necessarily rely upon PCAC to express the 
$\pi^\circ \rightarrow
\gamma\gamma$ amplitude in terms of the divergence of an AVV triangle.  
For such models,
however, one can no longer disregard quark masses.  The contribution to 
this amplitude of the
other side of the Ward-identity

\begin{equation}
[f_\pi m_\pi^2 \pi^\circ = ] \, \, \partial_\mu J_5^{3 \mu} = \alpha 
\tilde{F} \cdot F / 2 \pi +
2m_u \overline{u} \gamma_5 u - 2m_d \overline{d} \gamma_5 d
\end{equation}
{\it cancels} for sufficiently large quark masses $(m_{u,d} / m_\pi >> 
1)$, because of the
equivalence of the anomaly-term insertion and the PVV loop in the soft-
pion ({\it i.e.} large-
fermion-mass) limit, thereby confirming the Sutherland-Veltman theorem.  
The only way to make
such a model work is to introduce an ad hoc modification of PCAC itself 
through a redefinition
of the interpolating pion field: 

\begin{equation}
f_\pi m_\pi^2 \pi^\circ = \partial_\mu J_5^{3 \mu} - \alpha \tilde{F} 
\cdot F / 2 \pi ,
\end{equation}
Such a redefinition of PCAC [30,35] will ensure the usual anomaly-driven 
result in the
(constituent-quark-) limit of large quark masses for which the Sutherland-
Veltman theorem
applies--essentially by construction. One must therefore {\it alter} PCAC 
in order to obtain the
anticipated anomaly-driven result.  By performing this alteration, one is 
allowing the 
$\pi^\circ \rightarrow \gamma\gamma$ rate to determine PCAC, rather than 
the other way
around.

      Of course, it is certainly true that effective theories and the 
theories from which they are
derived must eventually yield the same results. It is certainly {\it 
possible} that the correct low-
energy approximation to QCD is described by a constituent-quark model with 
an appropriately
redefined
PCAC---to demonstrate this would entail the derivation of such a model 
from the QCD
lagrangian itself,
including a careful treatment of the appropriate anomaly functional
[36].  In the absence of
such a derivation, however, the redefinition (36) of PCAC constitutes a 
methodological {\it
prescription} for obtaining the correct answer with constituent-mass 
quarks, an answer that
emerges much more convincingly within a linear sigma model context
. \footnote{The usual operator statement of PCAC,
$\partial A^i = f_\pi m_\pi^2 \phi_\pi^i$, follows directly from the
linear sigma model lagrangian (e.g. ref [15]).}

        Such an effective theory is necessarily incomplete in the
absence of an explicit mechanism for confinement. When comparing a simple 
model 
such as the linear sigma model
with QCD in the infrared region, it is important to note that QCD is
a theory characterizing a number of nonperturbative phenomena,
including both confinement and chiral symmetry breaking.  In using the 
linear
sigma model as an approximation to low-energy QCD, we are necessarily
ignoring confinement in order to use chiral symmetry breaking to extract 
some of the dynamics of
light hadrons.  As is pointed out in a recent monograph on symmetry
breaking [37], such an approach hinges upon the assumption that the
dynamics responsible for these light hadrons is insensitive to
confinement -- in particular, that the size of the pion is smaller
than the hadronic confinement radius.  Implicit in this picture is
the idea of a distinct chiral-symmetry breaking momentum scale
$\Lambda_\chi (\simeq 1GeV)$ substantially larger than the
confinement scale $\Lambda_{QCD}$ [38] characterizing the formation
of massive hadrons.

\smallskip

\section * {6 Summary}

      In the present paper, we have considered the quark mass, the quark 
antiquark condensate,
and coupling strength $\alpha_s$ that emerge from the infrared region of 
QCD, and we have
argued the plausibility of identifying these quantities with those 
characterizing the linear sigma
model.  In particular, we have suggested that the quark mass and 
condensate characterizing the
linear sigma model be identified with the infrared freeze-out values of 
the dynamical mass and
condensate of QCD.  We also compare mass-gap tadpole equations in the 
linear sigma model and
in QCD, respectively, in order to associate $(g_{\sigma qq})^2 / 4 \pi$ 
with $\alpha_s^{eff}$,
the critical coupling characterizing chiral-symmetry breakdown in QCD. 
Independent
determinations of $\alpha_s^{eff} = \pi / 3$ and of $g_{\sigma qq} = 2 \pi 
/ \sqrt{3}$ appear
to corroborate these conclusions.

\smallskip

\section*{Acknowledgements}

Authors LRB and MDS are grateful for partial support from the US
Department of Energy.  VE would like to thank the Department of
Physics, University of Arizona, for its hospitality, and the Natural
Sciences and Engineering Research Council of Canada for research
support.

\eject

\newpage

\section*{Figures}

\smallskip
\begin{description}
\item{}Fig. 1:  The quark-antiquark condensate's equivalence to the trace 
of
the full fermion propagator loop [Eq. (6)] including a dynamical mass
function.

\item{}Fig. 2:  The quark-antiquark condensate's contribution to the quark
self-energy.  The exchanged particle is an SU(N) gluon.

\item{}Fig. 3:  L$\sigma$M representation of the constituent quark loop
origin of the pion decay constant $f_\pi$.

\item{}Fig. 4:  Contribution of the L$\sigma$M quark-antiquark condensate 
to
the L$\sigma$M quark mass.
\end{description}

\end{document}